\documentstyle[12pt]{article}
\title{Two-component scenario and related gaps in cuprates}
\author{N.Kristoffel\footnote{E-mail: kolja@fi.tartu.ee}
and P.Rubin\\
$\sp 1$Institute of Physics, University of Tartu,
Riia 142, 51014 Tartu, Estonia}
\date{}
\begin{document}
\maketitle
mechanism\\
Cuprate interband model\\

\begin{minipage}{13cm}
{\bf Abstract.} A simple model of cuprate superconductivity with an electron spectrum
prepared by doping is developed. The pair-transfer interaction couples the
itinerant band with two components ("hot'' and "cold'') of the defect
subsystem. There are basic defect-itinerant gaps quenched by progressive
doping. Band overlaps appear as novel sources for critical doping
concentrations. Insulator to metal transitions in the normal state are
expected here. Minimal quasiparticle excitation energies determine the
pseudo- and superconducting gaps according to the doping-dependent
disposition of bands. Two pseudogaps can be present at underdoping and two
superconducting gaps can be manifested at overdoping. Various
transformations and connections between the gaps agree qualitatively with versatile experimental
findings. The superconducting density does not reflect the presence of
"extrinsic'' gaps because of the interband nature of the pairing. A Uemura type
sublinear plot at underdoping with further recession is obtained. A mixed
Fermi-liquid is restored near optimal doping where the chemical potential
intersects all the band components. The metallization of the "cold''
subsystem is essential for the rise of T$_c$ on passing to optimal dopings.
\end{minipage}

\section{Introduction}
Extensive experimental data on superconductivity energetic characteristics have
been collected for cuprates [1-14].
However, standardized measurements
which cover self-consistently and smoothly the whole energy and doping
scales are advisable. The presence of at least one superconducting gap and
 one or even of two pseudogaps [5-7], rises a question
about the number of superconductivity order parameters. Indications of
the appearance of two superconducting gaps can be found. The nature of
the pseudogap which survives in the normal state has still remained debatable
\cite{15}. "Extrinsic'' and "intrinsic'' mechanisms have been proposed.
In the latter case the pseudogap is caused by the fluctuating
superconductivity order (preformed pairs) at $T>T_c$ and there is an
immediate connection between the superconducting gap and the pseudogap.
The extrinsic mechanism looks for the pseudogap source in the normal state
electron spectrum. Bare gaps in it may be due by other type orderings
(in lattice-phonon subsystem) or doping. At present such a type of
explanation seems to be preferable \cite{2}.

A cuprate superconductor can be considered as a charge-transfer insulator
perturbed by doping. In what follows the excitations in the charge
channel will be considered. The excitations in the spin-subsystem build
up an essential associated partner. Controversial statements on
interrelations of cuprate gaps have been made. These concern
transformations of superconducting gaps into pseudogaps on doping, the
connection of superconducting and normal state gaps and the coexistence of
various gaps in distinct doping regions. The authors believe that one
way to reach a more physical insight into cuprate superconducting properties
on the excitation-energy-doping phase diagram will be the elaboration of
a simple (possibly partly postulative) model by using the general
knowledge on these systems. The following comparison of the qualitative
outcome with observations can then illuminate the background physics.

Cuprate superconductivity is widely discussed in the two-component
scenario [16-18]. Its essence consists in the statement that a
"defect-polaronic'' subsystem bearing doped holes is functioning besides
the itinerant valence band electrons. The framework of the two-component
scenario leaves freedom for the precision of the nature of the 
electronic background and
pairing channels. One can e.g. consider as basic ingredients a mainly
oxygen band between the Cu-dominated Hubbard components and a distribution
of states created by doping [7,13,19-23] near the top of this band.
Lattice effects enter
this scenario through the inhomogeneous structure of CuO$_2$ planes in
doped cuprates (stripes, tweed patterns, granularity) and the associated
electronic phase separation [18,23-27].

Attempts to reflect the two-component scenario by a simple model have
been started in [19,20,28-30]. The present contribution is a
generalization, especially what concerns the superfluid density. Our model
uses a nonrigid electronic playground of superconductivity prepared by
doping. The "extrinsic'' source of the pseudogaps lies in the bare
normal state gaps between the defect and the itinerant states. 
These will be quenched with
progressive doping.
The mutual transformations between superconducting gaps and pseudogaps can
be explained by the change of the nature of the minimal quasiparticle
excitation energy on doping. The connection between pseudogaps and
normal state gaps is due to the mixing of the normal state spectrum with
the superconducting gaps in the quasiparticle energy expression. The
fluctuation effects have not been taken into account in this model, cf. \cite{15}.

The electron spectrum considered in the two-component scenario is nonrigid.
The appearance of a "defect'' subsystem besides the itinerant one opens a
novel channel for reaching high $T_c$-s by the interband pair transfer
interaction \cite{19,20}. The corresponding two-band superconductivity
mechanism \cite{31,32} has been known for a long time. A number of attempts
\cite{33} have been made to use it for cuprates in connection with the
two-component scenario [19,20,26,28-30]. The interband pairing interaction
operates in a considerable volume of the momentum space and works 
for pairing also as
being repulsive. It also prevents the manifestation of normal state
gaps in the superfluid density (order parameters) \cite{19,20}.

\section{The Model}
A cuprate electron spectrum created and reorganized by doping is described
as follows. The itinerant (valence) band ($\gamma$) states are lying
between the energies $\xi = -D$ and $\xi (max)=0$ and are normalized to $1-c$.
Here $c$ is a measure of the doped hole concentration. It must be scaled
for a given case, e.g. by joining  characteristic concentrations on the
phase diagram.

The defect subsystem is structurally anisotropic and this is manifested
in different gap features over the momentum space [1-3,13,40]. The
presence of two pseudogaps [5-7] of different behaviour is impressive. The
well-expressed large pseudogap is connected with the neighbourhood of the
"hot'' $(\pi ,0)$-type Brillouin zone points. The spectrum at
$(\frac{\pi}{2},\frac{\pi}{2})$-type "cold'' points seems to be weakly gapped \cite{40}.
For these reasons the defect system will be characterized by two subbands
for the "hot'' ($\alpha$) and "cold'' ($\beta$) regions, cf \cite{36}.
These subbands occupy the energy intervals $d_1-\alpha c$ and $d_2-\beta c$,
respectively, with the weight of states $c/2$. On underdoping these bands
lie above the valence band. Note that the optical charge-transfer gap is
reduced by doping \cite{41}. A progressive doping brings first the $\beta$-band
to overlap with the $\gamma$-band at $c_{\beta}=d_2\beta\sp{-1}$. A
common overlapping distribution of all the bands starts at
$c_{\alpha}=d_1\alpha\sp{-1}$. It is known that the infrared manifestation
of the defect band is lost at larger dopings in favour of a Drude peak
of (free) carriers \cite{42}.

The 2D (CuO$_2$ planes) densities of states of these bands read:
$\rho_{\gamma}=(1-c)D\sp{-1}$; $\rho_{\alpha}=(2\alpha )\sp{-1}$;
$\rho_{\beta}=(2\beta )\sp{-1}$ and $\beta <\alpha$ is supposed. There are
three qualitatively different arrangements of the bands and the chemical
potential ($\mu$). At $c<c_{\beta}$ $\mu =d_2-\beta c$ remains connected
with the "cold'' $\beta$-band. On underdoping the charge carriers are
concentrated in this "cold'' subsystem in accordance with \cite{13}. For
$c>c_{\beta}$, $\mu =(d_2-\beta c)[1+2\beta (1-c)D\sp{-1}]\sp{-1}$
intersects both ($\beta ,\gamma$)-bands. For the expressed dopings larger than
$c_0$, determined by $d_1-\alpha c_0=\mu$, the role of the ($\pi ,0$)-type
region increases essentially, cf. \cite{43,44}. Now the chemical potential
$\mu =[\alpha d_2+\beta d_1-2\alpha\beta c][\alpha +\beta +(1-c)2\alpha
\beta D\sp{-1}]\sp{-1}$ intersects all the three overlapping bands.

The valence band is attributed to the hole-poor regions of the material. It
remains the source of antiferromagnetic fluctuations whereas in the defect
space distinct spin-structures can be built up (ferrons, polaron aggregates,
etc.). Such type background has been used to describe the underdoped cuprate
magnetic properties \cite{37}. Presumably the defect part of the spectrum
can be compared with the bosonic (bipolaronic) component of the theories
including bozonization \cite{37,38}.

\section{Basic Expressions}
The cuprate pairing mechanism will be described by the coupling of
itinerant and defect subsystems through the pair transfer \cite{33}
interaction. Superconductivity is mutually induced in interacting components.
The corresponding coupling constant $W$ contains Coulombic and
electron-phonon (repulsive) contributions \cite{33}. Pairs are
formed from the particles of the same band.

The intraband contributions \cite{33} are of less significance on the present
level. The basic mean field Hamiltonian reads
\begin{equation}
H = \sum_{\sigma ,\vec{k},s}\epsilon (\vec{k})a\sp +
_{\sigma\vec{k}s}a_{\sigma\vec{k}s} +
\sum_{\vec{k}}\Delta_{\gamma}(\vec{k})[a_{\gamma\vec{k}\uparrow}
a_{\gamma -\vec{k}\downarrow}+a\sp +_{\gamma -\vec{k}\downarrow}
a\sp +_{\gamma \vec{k}\uparrow} -
\sum_{\vec{k},\tau}{}\sp{\tau}\Delta_{\tau}(\vec{k})[a_{\tau\vec{k}\uparrow}
a_{\tau -\vec{k}\downarrow}+a\sp +_{\tau -\vec{k}\downarrow}
a\sp +_{\tau \vec{k}\uparrow}]\; .
\end{equation}
Here $\epsilon_{\sigma}=\xi_{\sigma}-\mu$, $\sigma =\alpha ,\beta ,
\gamma$, $\tau =\alpha ,\beta$ and $\sum\sp{\tau}$ means the integration
with the densities of states $\rho_{\alpha ,\beta}$ in the corresponding
energy intervals. Usual designations are applicable
for spins ($s$) and electron operators.
The superconductivity order parameters are defined as
\begin{equation}
\begin{array}{cclcl}
\Delta_{\gamma}(\vec{q}) & = & 2\sum_{\vec{k},\tau}\sp{\tau}
W(\vec{q},\vec{k})<a_{\tau\vec{k}\uparrow}a_{\tau -\vec{k}\downarrow}>\\ \nonumber
\Delta_{\tau}(\vec{q}) & = & 2\sum_{\vec{k}}
W(\vec{q},\vec{k})<a_{\gamma -\vec{k}\downarrow}a_{\gamma\vec{k}\uparrow}>\; .
\end{array}
\end{equation}

The diagonalization of (1) yields the gap equation ($\Theta =k_BT$)
\begin{equation}
\begin{array}{cclcl}
\Delta_{\gamma}(\vec{q}) & = & \sum_{\vec{k},\tau}\sp{\tau}
W(\vec{q},\vec{k})\Delta_{\tau}(\vec{k})
E_{\tau}\sp{-1}(\vec{k})th\frac{E_{\tau}(\vec{k})}{2\Theta} \\ \nonumber
\Delta_{\tau}(\vec{q}) & = & \sum_{\vec{k}}
W(\vec{q},\vec{k})\Delta_{\gamma}(\vec{k})
E_{\gamma}\sp{-1}(\vec{k})th\frac{E_{\gamma}(\vec{k})}{2\Theta}
\end{array}
\end{equation}
with the usual form of the quasiparticle energies
\begin{equation}
E_{\sigma}(\vec{k})=\sqrt{\epsilon\sp 2_{\sigma}(\vec{k})+
\Delta\sp 2_{\sigma}(\vec{k})} \; .
\end{equation}
In what follows $W$ will be taken as constant. Moreover $\Delta_{\alpha}=
\Delta_{\beta}$ is set. At $T_c$, according to (3) the gaps
$\Delta_{\sigma}$ tend simultaneously to zero. For $W>0$ two
s-type order parameters appear with the opposite signs \cite{33}; expr. (1)
uses positive $\Delta$-s.

The number of paired carriers can be calculated as
\begin{equation}
n_s=\frac{1}{2}\left\{ \sum_{\vec{k}}\frac{\Delta_{\gamma}\sp 2(\vec{k})}
{E\sp 2_{\gamma}(\vec{k})}th\sp 2 \frac{E_{\gamma}(\vec{k})}{2\Theta}+
\sum_{\vec{k}}{}\sp{\tau}\frac{\Delta_{\tau}\sp 2(\vec{k})}
{E\sp 2_{\tau}(\vec{k})}th\sp 2 \frac{E_{\tau}(\vec{k})}{2\Theta}\right\} \; .
\end{equation}

Performing the integrations ($D, d>\Delta$) at zero temperature one finds \\
for $c<c_{\beta}$
\begin{equation}
n_{s0}=\frac{1}{2}\left\{ \Delta_{\alpha}\rho_{\beta}\arctan
\frac{\beta c}{\Delta_{\alpha}}+\Delta_{\gamma}\rho_{\gamma}
\left[ \frac{\pi}{2}-\arctan \frac{d_2-\beta c}{\Delta_{\gamma}}\right]
\right\} \; ,
\end{equation}
and for $c>c_{\beta}<c_0$
$$
n_{s0}=\frac{1}{2}\left\{ \Delta_{\alpha}\rho_{\beta}\left[ \frac{\pi}{2}-
\arctan\frac{d_2-\beta c-\mu}{\Delta_{\alpha}}\right]
+\Delta_{\alpha}\rho_{\alpha}\left[\frac{\pi}{2}-\arctan
\frac{d_1-\alpha c-\mu}{\Delta_{\alpha}}\right]+\right.
$$
\begin{equation}
\left.+\Delta_{\gamma}\rho_{\gamma}
\left[ \frac{\pi}{2}-\arctan \frac{\mu}{\Delta_{\gamma}}\right]
 \right\} \; .
\end{equation}
It can be seen that the presence of the normal state gap $d_2-\beta c$ does not
prepare a fermionic gap in the superfluid density. At the critical
concentration $c=c_{\beta}$ $n_s$ remains continuous (the second term in (7)
tends to zero).

\section{The Gaps}
Minimal quasiparticle energies reflect the presence of gaps in the
excitation spectrum of the superconductor. This will be the case also for
the pseudogaps, which appear naturally in the present model. The basic
source lies here in the perturbative segregation of the fermionic subsystem
by the presence and localization of the doped holes.

In the low underdoped region for $c<c_{\beta}$ one has
\begin{eqnarray}
E_{\alpha}(min)=\Delta_l=\sqrt{(d_1-\alpha c-\mu\sp 2)+
\Delta_{\alpha}\sp 2} \nonumber \\
E_{\beta}(min)=\Delta_{\alpha} \\
E_{\gamma}(min)=\Delta_s=\sqrt{\mu\sp 2+\Delta_{\gamma}\sp 2}\; . \nonumber \\
\end{eqnarray}

Further for $c>c_{\beta}$ and $c<c_0$ $E_{\gamma}(min)$ will be represented
by $\Delta_{\gamma}$ as $\xi_{\gamma}=\mu$ can be satisfied.
For $c>c_0$ also $E_{\alpha}(min)$ transforms to $\Delta_{\alpha}$.

In the normal state $\Delta_l$ and $\Delta_s$ survive
and one interpretes these as pseudogaps. A pseudogap and the
corresponding normal state gap are connected through the contribution of
$\Delta_{\alpha ,\gamma}$ into $E_{\sigma}$.
At $T=T_c$, $\Delta_{\alpha ,\gamma}$ vanish
and the quasiparticle nature of the excitation is lost. Passing to the
optimal doping the small pseudogap is smoothly transformed into the
itinerant superconducting gap. The large pseudogap regime extends until
$c\geq c_0$ is reached. The minimal excitation of the defect subsystem is
further determined by the superconducting gap $\Delta_{\alpha}$. The
connection and mutual transformation of the pseudogap and the corresponding
superconducting gap is based on the doping-variable structure of the
spectrum which changes the nature of the minimal excitation energy of
quasiparticles.

Concerning the manifestation of various gaps involved in the model, the
valence band excitations belong to the "hot'' spectrum. The "cold''
spectrum is usually considered as nongapped [1-3,40].
If one accounts for the
d-wave symmetry \cite{40} by multiplying $\Delta_{\alpha}$
with the corresponding
symmetry function, the "cold'' spectrum becomes empty. However, note that,
depending on the doping level and temperature, the two-band model allowes
pure $d$-, $s$- or mixed $d-s$ ordering symmetries \cite{45}. Extreme
dopings and low temperatures favour the $s$-wave nature. In spite of that
the basic gap manifestations appear in the "hot'' spectrum, the "cold''
electrons act essentially in building up the superconductivity high $T_c$,
supporting the interband pairing channel.

In summary the present model predicts the appearance of two pseudogaps on
low dopings. Further the spectrum involves the large pseudogap and the
itinerant superconducting gap. On overdoping the spectrum is expected to
contain two superconducting gaps. Then the defect
$\Delta_{\alpha}$ will be manifested by
an additional spectral weight inside of $\Delta_{\gamma}$. In the case
when the $\beta$-subsystem states overlap the itinerant band from the very
beginning there will be only one (defect subsystem) pseudogap.

\section{The Illustrations and Comparative Discussion}
The theoretical cuprate phase diagram following from the present model is
illustrated by Figures 1 and 2. The following plausible parameter set has
been used: $D=2$; $d_1=0.3$; $d_2=0.1$; $\alpha =0.66$; $\beta =0.33$ and
$W=0.28$ (eV). At this $T_c(max)=125$ K is reached for $c=0.57$ and
$c_{\alpha}=0.45$; $c_{\beta}=0.38$; $c_0=0.57$. The scaling for a typical
cuprate doping is made by $p=0.28c$ according to the widely accepted
value $p=0.16$ corresponding to $T_c(max)$. The gaps in Fig.1 are given
for $T=0$ and the connections between them can be followed on the doping
scale. There seems to be a general agreement with the findings for cuprates.

The expected common manifestation of two underdoped state pseudogaps has
been established for the La- and Bi-cuprates [5-7].
Another class of compounds with one charge channel pseudogap (bare
nongapped $\beta$-subsystem) with the $\Delta_l$-type behaviour
is eventually possible. The smooth
transformation of the small pseudogap into the larger superconducting gap
(at $p_{\beta}$) has also been observed \cite{5,6,13}. The large pseudogap
extends to slight overdoping and then transforms into the defect system
superconducting gap as has been found in Ref.[46]. $\Delta_l$ is attributed to
the spectral hump-feature \cite{9,10}. On intermediate dopings $\Delta_l$
and the itinerant superconducting gap appear together. They cross close
to the optimal doping. This corresponds to the observations \cite{9}. Note
that in a narrow doping region the larger superconducting gap exceeds
the pseudogap $\Delta_l$. The corresponding parameters of the itinerant
and defect subsystem are not competing. Eventually, the hump is shifted
to larger energies with reduced dopings as observed in \cite{10} and it
remains preserved for $T>T_c$ on the dopings where $T_c$ is optimized
\cite{9}.

Following the connection of the pseudogap with the "own'' subsystem
superconducting gap it can be seen that the manifestation of a superconducting
gap on a given doping can be substituted by the appearance of the normal
state gap (Fig.2) for $T>T_c$ in this region. It means that at low
temperatures a pseudogap can remain not manifested on dopings where it
will be found in the normal state (cf. \cite{14}). In general the pseudogaps
persist to $T=0$. The rising temperature expands slightly the manifestation
region of a pseudogap on the doping scale.

The manifestation of both superconducting gaps on overdoping is often
debated. However the Fermi energy intersects the electron spectrum parts
headed by different band components at different wave vectors and the
larger $\Delta_{\gamma}$ can remain masked.

The temperature dependence of the pseudogaps remains at the present state
of the model due by the contribution of the superconducting gaps. This
leads to a slow diminishing with $T\rightarrow T_c$, whereas the
superconducting gaps $\Delta_{\alpha ,\gamma}$ reach zero at $T_c$ in a
"traditional'' manner, cf. \cite{47}. Figure 2 represents the behaviour
of the superconducting density ($T=0$) together with the normal state
gaps and $T_c$. There are no signs of bare "extrinsic'' 
fermionic gaps in a continuous
curse of $n_s$ and an argument \cite{15} against the "extrinsic'' source
of the pseudogap fails. This is the result of the interband nature of the
pairing. The normal- and pseudogaps and their large ratios to $T_c$ rise
dramatically with the diminishing of doping, whereby the superconducting gaps and
density decrease.

In Fig.3 the calculated Uemura type \cite{48} plot is shown. On
underdoping there is a sublinear segment connecting $T_c$ and $n_s$ with
the further recession.

The dynamics of the band overlap in the present model introduces a novel
source for special critical dopings. These correspond to the doping
concentrations where the band components begin to overlap. The metallization
of the cold defect-liquid is reached at $c_{\beta}$. On smaller dopings
one supposes the formation of doped hole ferrons and a percolation type
superconductivity \cite{49}. The defect band acts as a bath of
uncompensated spins. One can explain \cite{37} the presence of a magnetic
pseudogap \cite{50,51} in the spin excitation channel on such basis.

The hot defect subsystem metallizes at $c_0$ near the optimal doping. Here
all the three bands of the model overlap and are intersected by the chemical
potential. This means the built up of a common mixed Fermi liquid. The
difference between the defect and the itinerant carriers is washed up. The large
pseudogap gets lost when passing this border. Here one expects an insulator to
metal transition in the normal phase.
During the way to this concentration the Fermi surface becomes
more and more electron like with
peripheral hole pockets. Experiments on the normal phase \cite{52} show
that a quantum metal to insulator transition appears at a distinct $c_k$ in
the same region as $c_0$ lies. For smaller dopings the hot quasiparticles
become insulating where the cold quasiparticles remain metallic. Various
experimental findings add to the existence of a critical doping
concentration in cuprates \cite{21,53,54}, where the properties of the electron
liquid are essentially changed. Supposing that $c_k=c_0$,
these findings become
qualitatively explained. Such a $c_k$ is of a basic importance in a quantum
critical point scenario \cite{53}.

Some further essential properties of cuprates can be relatively simply
explained by two-band models. The two observed electronic relaxation
channels \cite{46} and coherence lengths are a natural property of two-band
superconductors \cite{55}.

The transition temperature and effective mass isotope effects can also be
explained in two-band schemes \cite{56,57}. In general one observes a weak
transition temperature effect for its optimal values and vice versa. This
behaviour is caused by a contribution of a repulsive electron-phonon
interaction in the whole pair-transfer scattering. This contribution of
some percent in magnitude can cause the observed $T_c$-shifts. The pseudogap
isotope effect \cite{58} seems to be more complicated, being connected
with the changes of both components in the quasiparticle energy.\\

The present simple model with plausability elements seems to be able to
reproduce qualitatively the behaviour of energetic characteristics of
cuprate superconductors. There remains a wide freedom to fill it in with
better substantiated suppositions and quantitative aspects.\\

This work was supported by Estonian Science Foundation grant No 4961.

\newpage
Figure captions\\

Fig.1. Doping dependences of gaps: 1 -- the large pseudogap $\Delta_l$;
2 -- the small pseudogap $\Delta_s$; 3 -- the itinerant system
superconducting gap $\Delta_{\gamma}$; 4 -- the defect system
           superconducting gap $\Delta_{\alpha}$; 5 -- $T_c$. $p = 0.28c$;
$p_{\alpha}=0.13$; $p_{\beta}=0.085$; $p_0=0.16$; $p(T_{cm})=0.16$.\\

Fig.2. Doping dependences of the superconducting density (curve 1) and
of the transition temperature (curve 2). Curves 3 and 4 represent the normal
state gaps.Energetic characteristics are given in eV. \\

Fig.3. Transition temperature vs. the superconducting density
(the Uemura plot).

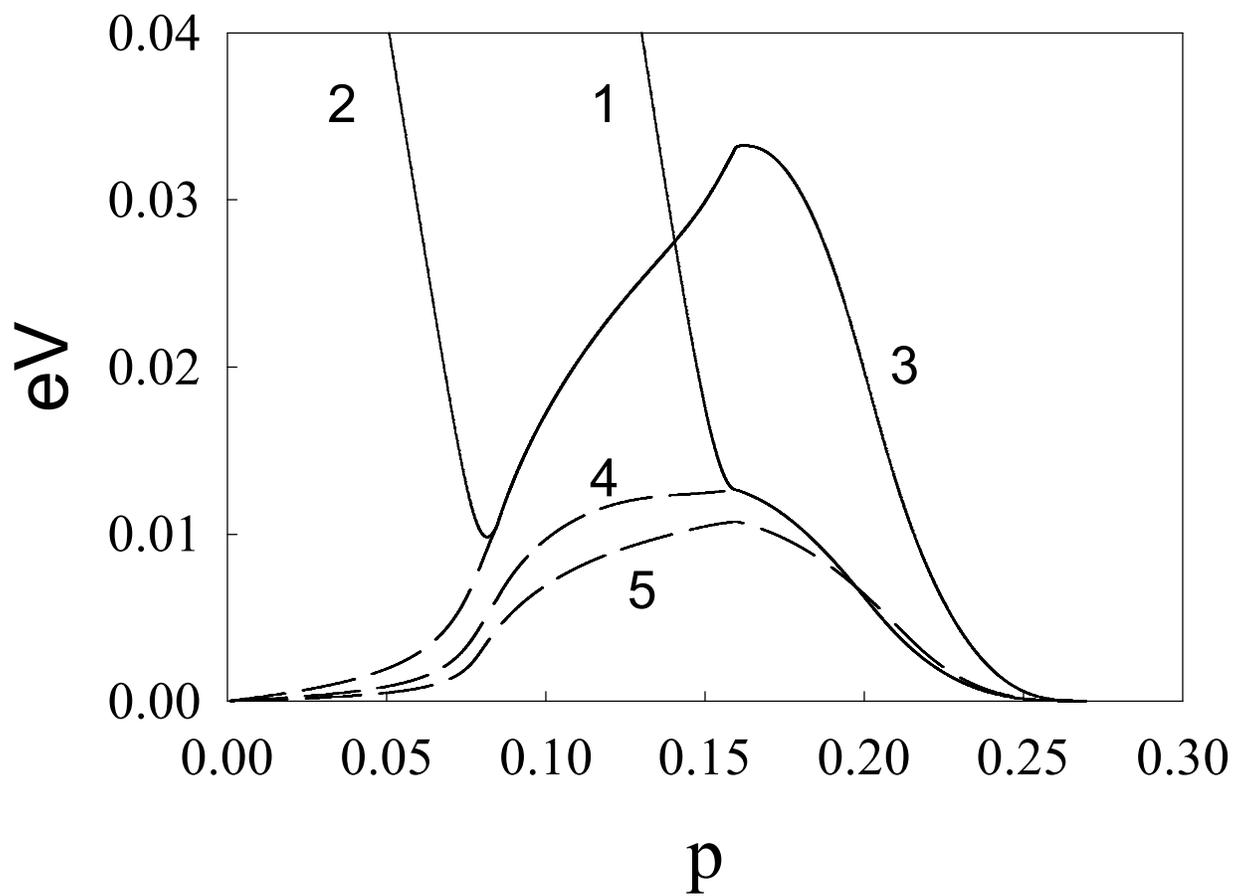
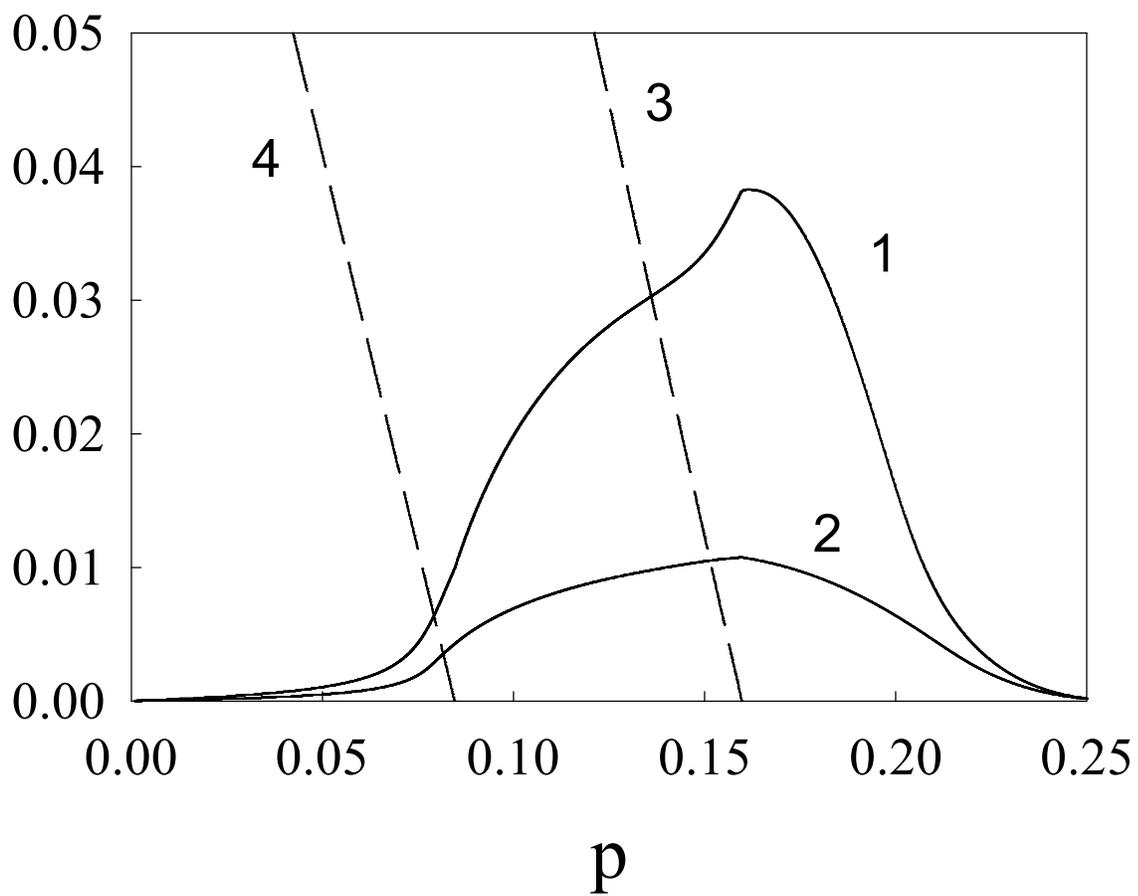
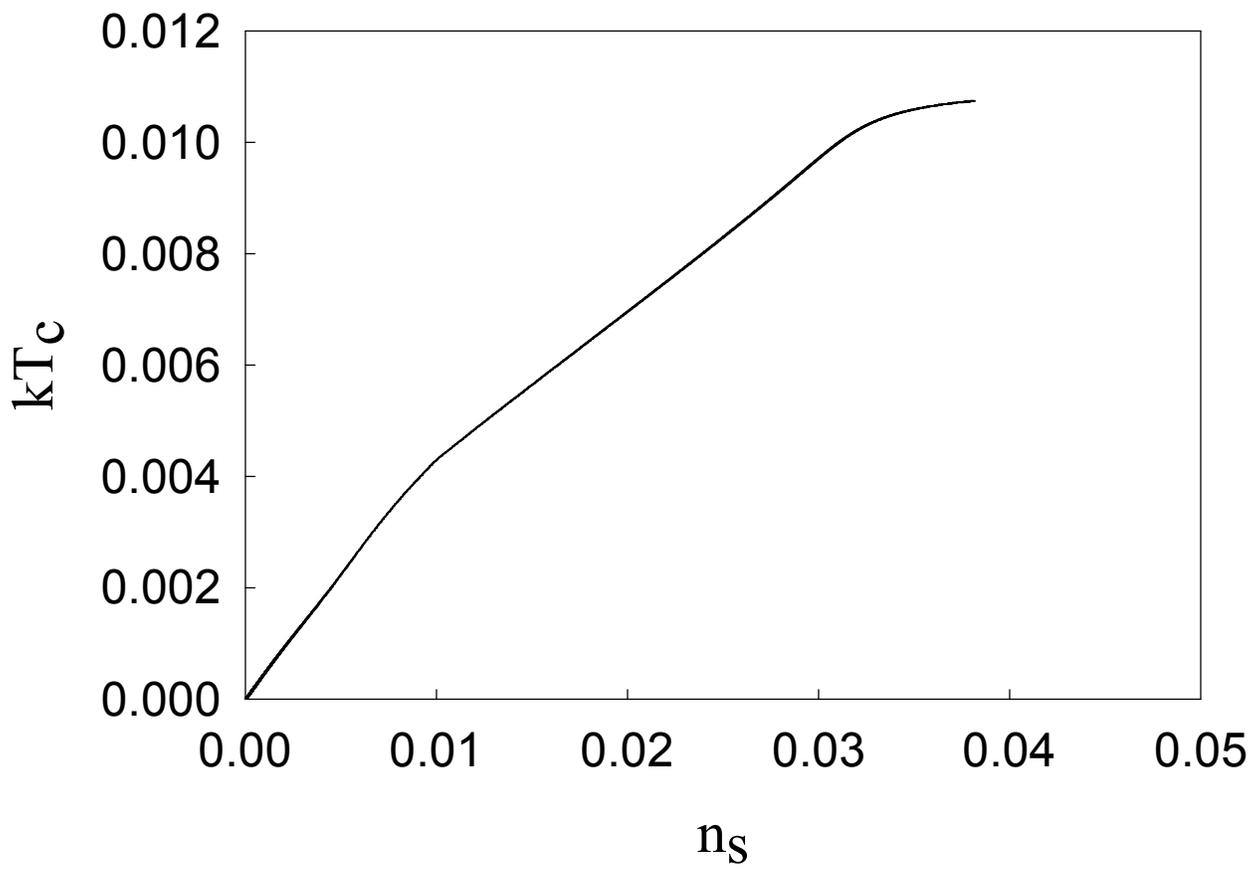
\end{document}